\documentclass[preprint]{aastex}

\begin{document}

\title{The first magnetic fields in the universe} 

\author{Andrei Gruzinov}

\affil{CCPP, Physics Department, New York University, 4 Washington Place, New York, NY 10003}

\begin{abstract}

We show that the first structures that form in the universe should spontaneously generate magnetic fields. No primordial seed field is required for this ``first dynamo''.

Although the first dynamo starts with kinetic plasma instabilities, we argue that an adequate magnetohydrodynamic description might be possible via a simple trick. This should allow a numerical study of the effect of the first magnetic fields on the first baryonic objects.

\end{abstract}

\section{Introduction}

It has already been proposed that the very first objects that form in the universe should generate magnetic fields (Gruzinov 2001, Rees 2006). Here we offer a concrete scenario. In our scenario the first magnetic fields appear before the first stars, and the dark age ends with the synchrotron light from the first shocks. This light is probably unobservable. But the first magnetic fields can, in principle, change the first star or black hole formation. 

Consider a naive version of the magnetic field evolution at high redshift, $z\sim 100$, when the universe contains no stars or quasars, and the very first baryonic structures begin to form. The gas turbulence generated by the structure formation amplifies the magnetic field, because the gas is a good conductor. The growth time of the field is of order the dynamical time, and the first structures amplify the primordial magnetic field by a factor of a few. Since the primordial field must be very weak \footnote{ For all we know, there might be no primordial magnetic field at all. There are some speculative theories of primordial magnetogenesis, but they too predict only a weak field.}, we must conclude that the first structure formation is virtually unaffected by the magnetic field.

This logic is flawed, because the exponentiation time of the magnetic energy is actually much shorter than the dynamical time. We explain the main idea in \S2, give the full scenario in \S3, where we also provide the relevant numerical estimates. In \S4 we propose a simple method which should allow to use numerical magnetohydrodynamics to study the first magnetic fields.

\section{The main idea}

Our full scenario is too complex and multi-stage to be trusted in detail. But the overall result -- very roughly equipartition large-scale magnetic field in a few dynamical times -- hinges on just one simple observation. Namely, we note that the number of eddy turnover times at the small-scale end of the Kolmogorov cascade is $N\gg 1$. For all practical purposes, $e^N=\infty$, leading to an equipartition field even starting from a zero seed.

To be concrete, consider a merger of two gas clouds of size $\sim R$. The merger shocks and stirs the gas for a few dynamical times $t_d\sim {R\over V}$, where $V$ is the velocity of the merger. The resulting gas motion is hydrodynamic: $R$ is much larger than the mean free path of the hydrogen atom $\lambda$ (this allows to simulate the structure formation using hydrodynamics). 

The gas thermal velocity should be of order $V$, because numerical experiments show significant heating during the first mergers (Abel et al 1998). Then the gas viscosity is $\nu \sim V\lambda$.  Since the  Reynolds number is $Re\equiv {VR\over \nu }\sim {R\over \lambda}\gg 1$, the gas motion is turbulent. 

The Kolmogorov  cascade to small scales develops. The characteristic velocity of eddies of size $l$ is $V_l\sim V({l\over R})^{1/3}$. The eddy turnover time at scale $l$ is $\tau _l\sim t_d ({l\over R})^{2/3}$. The Kolmogorov cascade cuts off when viscosity starts to dominate, $\nu l^{-2}\sim \tau _l^{-1}$, giving the cutoff scale $l_\nu \sim ({R\over \lambda })^{1/4}\lambda$. The number of turns of the smallest eddy in the cascade is $N\sim {t_d\over \tau _{l_\nu }}\sim ({R\over \lambda })^{1/2}$. 

For example, with $R\sim 10^{20}$cm and $\lambda \sim 10^{14}$cm \footnote{Here and below all length scales are physical, with the numbers loosely taken from Abel et al (1998).}, we get $N\sim 1000$. If the small-scale magnetic field exponentiates after each turn of the smallest eddy, the growth factor is $\sim e^N\sim 10^{400}$. This number is clearly too big to mean anything. All it says, is that an arbitrarily weak small scale field will saturate in a time $\ll t_d$.

Then the magnetic field will climb up the Kolmogorov cascade, staying in rough equipartition with the eddies of the covered scales. The time to climb the entire cascade is dominated by the last step, because $t_d+{t_d\over 2^{2/3}}+{t_d\over 4^{2/3}}+...\sim t_d$.

\section{The scenario}

As we said, with the amplification factors $\sim 10^{400}$, it does not really matter where the seed field is coming from. But we think the full story develops roughly by the following scheme:

shot noise at Debye length $\delta$ --- kinetic plasma  instabilities (scales above $\delta$)  --- small-scale dynamo (scales below $l_\nu$)  --- dynamo (scales above $l_\nu$, climbing up the Kolmogorov cascade).

\subsection{Debye length and kinetic plasma instabilities}

The Debye length $\delta$ and the plasma frequency $\omega_p$ are defined as $\delta \equiv ({T\over 4\pi x_ene^2})^{1/2}\sim {V\over \omega _p}$, where $T$ is the gas temperature, $x_e$ is the ionization fraction, $n$ is the gas density. With $T\sim 1000$K, $x_e\sim 10^{-4}$, and $n\sim 1$cm$^{-3}$, we get the thermal velocity $V=3\times 10^5$cm/s, the Debye length $\delta \sim 3\times 10^4$cm, and the plasma frequency $\omega _p\sim 10$s$^{-1}$.

The mean free path of the hydrogen atom is $\lambda \sim {1\over n\sigma }\sim 10^{14}$cm, where $\sigma \sim 10^{-14}$cm$^2$ is the atomic hydrogen scattering cross section. The mean free path of charged particles from Coulomb scattering $\lambda _c\sim {10^4T^2\over x_en}\sim 10^{14}$cm is of the same order. 

Since $\delta \ll \lambda $, the shocks in the merging gas develop kinetic instabilities. The growth rate of kinetic instabilities is $\sim \omega _p$, the length scales are $\sim \delta$. The seed for these instabilities is provided by the shot noise at the Debye scale: $(x_en\delta ^3)^{-1/2}\sim 10^{-5}$. This is a sufficient seed, because the number of e-foldings of kinetic instabilities can be estimated as $\omega _p{\lambda \over V}\sim 3\times 10^9$. The instabilities are mostly electrostatic, but for our purposes, it is enough to note that magnetic fields of order ${V\over c}$ will accompany any electric field in a moving plasma.

There is a kinetic instability (Weibel)  which gives magnetic fields directly (Sagdeev 1966), but we don't have to worry about these details here. We simply note, that due to kinetic instabilities, we get a shot-noise magnetic field on scales $l$, very roughly of order $({\delta \over l})^{-3/2}$.

\subsection{Magnetic Reynolds number and the small-scale dynamo}

The plasma conductivity is $\sigma \sim {x_ene^2\over m_e}{\lambda \over 40V}\sim 3\times 10^{11}$s$^{-1}$. This gives the magnetic diffusivity $\eta \equiv {c^2\over 4\pi \sigma}\sim 3\times 10^8$cm$^2$s$^{-1}$, and the magnetic Reynolds number $Rm\equiv{VR\over \eta}\sim 10^{17}$. 

For $Rm\gg Re$, the magnetic fields exists on small length scales, below the hydrodynamic turbulence cutoff scale $l_\nu$. On these length scale the hydrodynamic flow acts as a linear incompressible deformation. On the average, these linear deformations amplify the magnetic field \footnote{ Kulsrud, Andersen (1992) give a physical discussion, Gruzinov et al (1996) give a rigorous proof. }. 

The growth rate of the small-scale dynamo on the scales below $l_\nu$ is $\sim \tau _{l_\nu }^{-1}\sim Nt_d^{-1}$. The cutoff scale of the small-scale dynamo $l_m$ is set by the resistive damping, $Nt_d^{-1}\sim \eta l_m^{-2}$, giving $l_m\sim Re^{-1/4}Rm^{-1/2}R\sim 10^{10}$cm. Since the number of e-foldings of the small-scale dynamo is $N\sim 10^3$, the shot noise seed from the kinetic dynamo will suffice to saturate the magnetic field on scales below $l_\nu$ in a time $\ll t_d$.

\subsection{Climbing up the Kolmogorov cascade}

Assuming the small-scale dynamo provides a seed on the scale $\sim l_\nu$, what should we expect on scales $l>l_\nu$ in a turbulence with $Rm\gg Re \gg 1$? There is no real theory for this process \footnote{Because, by the ``3D Zeldovich theorem'', magnetic field changes the turbulence long before it reaches the large-scale equipartition (Gruzinov, Diamond 1994).}, but we expect that the magnetic field will climb up the Kolmogorov cascade, spreading to larger and larger scales. In a few dynamical times, a roughly equipartition, roughly large-scale field should appear.

This behavior is dictated by the following three  principles. First, magnetic energy is not conserved, and should spread to all available scales ``by ergodicity''. Second, the spread of magnetic energy should be approximately local in the wavenumber space. Third, the only reasonable saturation level of the field amplitude is the equipartition with the flow \footnote{ But we must note, that in a somewhat different context of relativistic dynamo, where we predicted ``very roughly equipartition large-scale magnetic field in a few dynamical times'', Zhang et al (2009) actually find that although the large-scale field indeed grows at a much faster rate than $t_d^{-1}$, after a few $t_d$ the large-scale magnetic field stays below the equipartition.}.

\section {Structure formation with magnetic field}

Assume that the first dynamo does operate. What are the consequences? The synchrotron emission of the first shocks is negligible. It appears that the first magnetic fields are potentially important for just one thing, the structure formation. In particular, they might change the initial mass function of the first stars and black holes.

Obviously one cannot directly simulate the kinetic instabilities and the structure formation in one go, because $\omega _p^{-1}\sim 0.1$s and $t_d\sim 3\times 10^{14}$s. One cannot even simulate the hydrodynamic part of the problem, because both $Re$ and $Rm$ are too large.

But we think there is a way to include the first magnetic fields in the structure formation simulation. One should run several simulations with $Rm > Re> 1$, where ``$>$'' means greater by as much as numerically possible. The simulations should include an initial seed field of tunable magnitude. If the overall scenario of \S3 is correct, one will find that the simulation results are roughly independent of the seed field magnitude, if the seed magnitude lies in a certain interval.

\acknowledgements 

This work was supported by the David and Lucile Packard foundation.

\end{document}